# Memory Metamaterials

Tom Driscoll[1], Hyun-Tak Kim[2], Byung-Gyu Chae[2], Bong-Jun Kim[2], Yong-Wook Lee[2, 3], Nan Marie Jokerst[4], Sabarni Palit[4], David R. Smith[4], Massimiliano Di Ventra[1], Dimitri N. Basov[1]

1. Physics Department University of California at San Diego, La Jolla, California 92093, USA

2. IT Convergence & Components Lab, ETRI, Daejeon 305-350, Republic of Korea

3. School of Electrical & Engineering, Pukyong National University, Busan 608-739, Republic of Korea

4. Center for Metamaterials and Integrated Plasmonics and Electrical and Computer Engineering Department, Duke University, P.O. Box 90291, Durham, North Carolina 27708, USA

**The resonant elements that grant metamaterials their unique properties have the fundamental limitation of restricting their useable frequency bandwidth. The development of frequency-agile metamaterials has helped to alleviate these bandwidth restrictions by allowing real-time tuning of the metamaterial frequency response. We demonstrate electrically-controlled persistent frequency tuning of a metamaterial, allowing lasting modification of its response using a transient stimulus. This work demonstrates a form of memory capacitance which interfaces metamaterials with a class of devices known collectively as memory devices.**

The ability of metamaterials to create electromagnetic responses absent in nature has initiated the new research field of transformation optics (1,2), which has applications ranging from electromagnetic cloaking (3) to sub-diffraction imaging (4). Frequency-agile metamaterials, which allow one to adjust the electromagnetic response in real time, are emerging as an important part of this field. The hybrid-metamaterial approach (5,6), wherein natural materials are integrated into the metamaterial composite, has been particularly successful in enabling frequency-agile metamaterials which respond to application of voltage (7,8), external electric field (9), light (10,11), and heat (12). This tuning ability helps make metamaterial devices more versatile, adapting to shifting input or changing target parameters. However, those methods for enabling frequency-agile metamaterials require continuous application of an external stimulus to maintain altered metamaterial properties. Once the external stimulus is removed, the metamaterial returns to its original response. Essentially, any functionality derived from metamaterials would benefit greatly if the metamaterial tuning persisted once the triggering stimulus disappeared. Metamaterials which are tuned mechanically or geometrically should retain their tuned properties (13,14), but such techniques are likely to be difficult to implement at higher frequencies or for complex designs. We have achieved an electrically-controlled memory effect using a hybrid device composed of a resonant metamaterial and a transition metal oxide. The principle underlying the persistent tuning of our hybrid device is that of memory-capacitance (memcapacitance for short) recently suggested theoretically (15).

The persistent frequency tuning of a memory metamaterial device can be illustrated using a single layer gold split-ring resonator (SRR) array patterned (16) on a 90nm thin film of vanadium dioxide ($VO_2$) (see Fig.1A). $VO_2$ is a correlated electron material that exhibits an insulator-to-metal (IMT) phase transition which can be thermally (17), electrically (9,18), or optically (19) controlled. The IMT is percolative in nature and is initiated by the formation of nanoscale metallic puddles in the insulating host (20). The transition is highly hysteretic, and has been previously shown to exhibit memory effects (21). The hysteresis associated with the $VO_2$ can be observed by measuring the DC resistance of the sample (Fig.1B solid lines). The phase transition also affects the dielectric properties of $VO_2$ in a unique way. At the onset of the IMT, electronic correlations acting in concert with the spatial inhomogeneity of $VO_2$, create a sharply divergent permittivity (12,20). This increasing $VO_2$ permittivity increases the capacitance of the SRR resonators such that the metamaterial resonance frequency decreases as the IMT progresses. The data in Fig.1B,C illustrate this effect. At room temperature, we identify the resonance frequency of the SRR array as the spectral minimum at $\omega_0$=1.65 Terahertz (THz) (darkest line in Fig.1C). As the dielectric constant of $VO_2$ is increased with temperature, the resonance frequency red shifts by as much as 20%. The discussion below shows that this metamaterial resonance tuning persists when accomplished via short current pulses.

Because both the DC resistance and the permittivity are products of the IMT, the SRR resonance in the hybrid-metamaterial inherits the same hysteretic nature observed in the DC resistance (21). Any transient excitation which causes a small perturbation in the $VO_2$ IMT will leave a lasting change of the resonance of the metamaterial. We can apply such an excitation voltage pulse using the electrical leads connected to the device. In this case, the applied voltage promotes the IMT by inducing local heating as current flows through the $VO_2$ film (22).

In order to maximize the observable effect small voltage pulses have on the metamaterial, we adjust the temperature of the device to a region where the hysteresis is most pronounced. This corresponds to a range where the slope of the resistance as a function of temperature, $R(T)$, is the steepest, and we choose 338.6 Kelvin (vertical dashed line in Fig.1B). Applying a series of voltage pulses (Fig.2B), while spectroscopically monitoring the resonance frequency of our hybrid structure before and after each pulse (12), we see the effect of these sequential voltage pulses (Fig.2A). Each pulse consistently red-shifts the resonance frequency. This red-shift is clearly visible even after the voltage is removed, demonstrating that a persistent change in the hybrid-metamaterial was obtained. It is important to note that the volumetric heat-capacity of the whole device is quite large given the rather limited power input with each pulse. Thus, this is not a global thermal effect; rather, heating is localized primarily to the $VO_2$ film, and is transient. After the voltage pulse subsides, the device rapidly thermalizes back to the starting temperature. Temperature monitoring confirms the overall temperature is unchanged to within ±0.02 Kelvin during the entire duration of the experiment. The persistence of the modified resonance has been spectroscopically monitored for 10 minutes and did not show signs of degradation. Our work on the DC memory resistance of $VO_2$ suggests the effect will persist much longer (25). (21).

It is instructive to analyze the response of our device within an effective circuit model (23,24). Our SRR array can be modeled as an array of RLC circuit elements each with resonance frequency $\omega = (LC)^{-1/2}$ (Fig.2C). The inductance **L** is known to remain constant, because no permeability changes occur in any material within our device. This allows us to relate changes of the observed resonant frequency directly to a variation in capacitance: $\frac{C}{C_0}=\left(\frac{\omega_0}{\omega}\right)^2$. We estimate the capacitance $\mathbf{C_0}$ of a single SRR of our dimensions as $C_0 \approx 2.5\text{x}10^{-15}$ Farad/SRR (25). The green points in Fig.2A show the SRR capacitance in this manner. The plotted capacitance **C** is the total capacitance of each SRR. To model the persistent-tuning behavior of our device within the effective circuit picture, we introduce the circuit element of memory capacitance $\mathbf{C_m}$ (15), defined by the relations $q(t)=C_m(x,V_C,t)V_C(t)$, $\dot{x}=f(x,V_C,t)$. $V_C$, $q(t)$ are the bias and charge, respectively, on the capacitor defined by $VO_2$, and $x$ is a set of state variables which, in the present case, accounts for the dielectric behavior of the $VO_2$ IMT. $\mathbf{C_m}$ acts in parallel with the SRR natural capacitance to give a total capacitance $\mathbf{C}=\mathbf{C_0}+\mathbf{C_m}$. In particular, $\mathbf{C_m}$ parameterizes the metamaterial-tuning memory effect demonstrated in Fig.2A. The increasing conductivity of $VO_2$ as the IMT progresses also introduces a memory resistance ($\mathbf{R_m}$) (15, 26 and references therein) (Fig.2B). $\mathbf{R_m}$ acts detrimentally to reduce the quality factor of the metamaterial resonance in our device. Thus softening of the resonance which accompanies the reduced resonance frequency is apparent in the spectra (Fig.1C). In $VO_2$ the memory resistance and memory capacitance appear to be inextricably intertwined; an effect that has also been anticipated theoretically, and which should be particularly relevant at the nanoscale (15). In our circuit model $\mathbf{V_{ext}}$ is the applied voltage pulse which modifies $\mathbf{C_m}$ and $\mathbf{R_m}$. $\mathbf{V_{IR}}$ originates from our infrared probe. Very different powers and timescales allow us to effectively separate the operation of the probing $\mathbf{V_{IR}}$ and modifying $\mathbf{V_{ext}}$ into the SRR and $VO_2$ dominated groups (Fig.2C). Specifically, the resonance frequency of the SRRs is much faster than the timescales of $\mathbf{V_{ext}}$, and the power levels used for spectroscopic probing are much less than those necessary for modification of **C**. Systems where these conditions are not true will make interesting nonlinear study cases.

To better illustrate the operation of the device, we propose a thermal finite-element model (16) to examine the time-evolution of the $VO_2$ temperature for a range of input powers (Fig.3A). The input power of the first four voltage pulses ($\alpha$,$\beta$,$\gamma$,$\delta$) are marked, and we see that for each a near steady-state temperature-rise (**ΔT**) quickly emerges. This thermal result enables a more detailed look at the evolution of capacitance during voltage pulses (Fig.3B). Starting from data point **A**, with normalized capacitance **C**=$\mathbf{C_0}$, the voltage pulse $\alpha$ raises the temperature of the $VO_2$ by $\mathbf{\Delta T_\alpha}$=0.11 Kelvin (as documented by the finite-element simulations), causing an increase in the capacitance. Once the voltage pulse stops, the $VO_2$ quickly (~25ms, (16)) thermalizes back to the bias temperature of 338.6 Kelvin. However, because of the hysteresis in the IMT a net change in the capacitance persists, and the device settles to point **B** with capacitance **C**=1.006 $\mathbf{C_0}$. These simulations reveal the complete heating/cooling cycle, which gives the switching timescale as set by the thermal geometry of the device close to ~50ms. The arrows in Fig.3B illustrate this hysteretic capacitance behavior by showing a probable path for **C(T)** during this heating/cooling cycle between data points. These arrows are drawn using the experimentally determined pre- and post- pulse capacitances from Fig.2A, and the maximum temperature rise revealed from simulations in Fig.3A. This is coupled with an assumption that the hysteresis in the IMT is strong enough that the post-pulse thermalization cooling has little effect on the capacitance value. This latter assumption is well-supported by the much flatter slope observed in Fig.1B for the DC resistance during cooling than heating at 338.6K ($dR_{heat}/dT$ = 100kΩ/°K. $dR_{cool}/dT$ = 4Ω/°K), as well as other work (21). However, with our current equipment configuration we are unable to experimentally probe the capacitance on fast enough time scales to directly obtain this **C(T)** data inside pulses. Nevertheless, it is clear that the demonstration of electrically-controlled memory capacitance in this device relies on the hysteretic nature of the $VO_2$ phase-transition.

It is finally worth stressing that the use of a temperature bias in our experiment is required only to put a particular $VO_2$ film into a regime where the IMT is highly hysteretic, although our $VO_2$ does exhibit some hysteretic qualities even at room temperature (16). More promisingly, several $VO_2$ fabrication techniques are known to reduce the phase-transition temperature down to room temperature (27,28) and thus will enable $VO_2$-hybrid memory metamaterials to operate at ambient conditions. Additionally, any material which posses a hysteretic response in either its permittivity or permeability at a suitable frequency range could be used in a hybrid-metamaterial design to obtain memory effects analogous to those we demonstrated here (29). The correlated electron state which gives $VO_2$ the divergent permittivity used in this demonstration is only effective up to mid-infrared. Combination of other hysteretic materials with metamaterials operational in near-infrared and visible could easily push this effect to higher frequencies which are beneficial for a variety of practical applications (30).

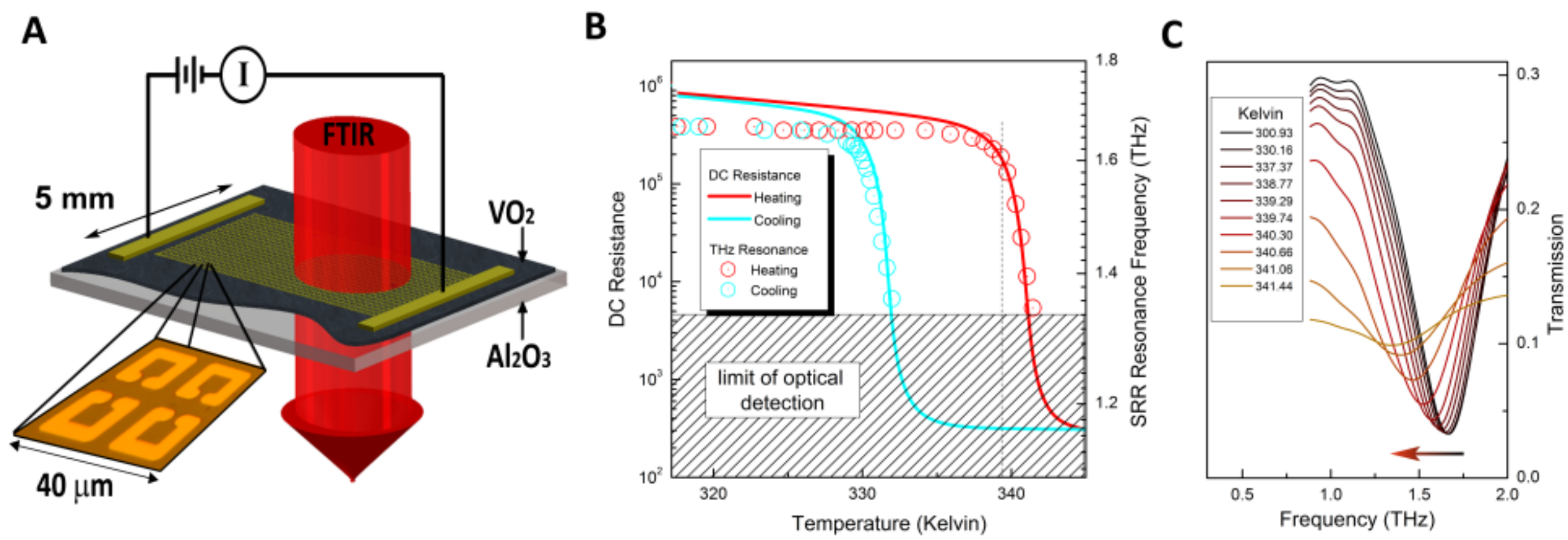

**Figure 1. Memory-oxide hybrid-metamaterial device. A**, The device consists of a gold split-ring resonator array which has been lithographically fabricated on a vanadium dioxide film. Electrodes are attached allowing in-plane I-V transport, and the device is mounted to a temperature control stage. **B**, Simultaneous DC-transport and Far-infrared probing of the metamaterial demonstrate that as $VO_2$ passes through its insulator-to-metal transition, resistance drops and the SRR resonance frequency decreases. This latter effect occurs due to an increase in the per-SRR capacitance. **C**, Spectroscopic data from which the metamaterial resonance frequencies in **B** are identified (heating cycle shown). Data in panels B and C are obtained by performing a complete temperature cycle (300K to 350K to 300K) using the temperature stage on which the sample is mounted.

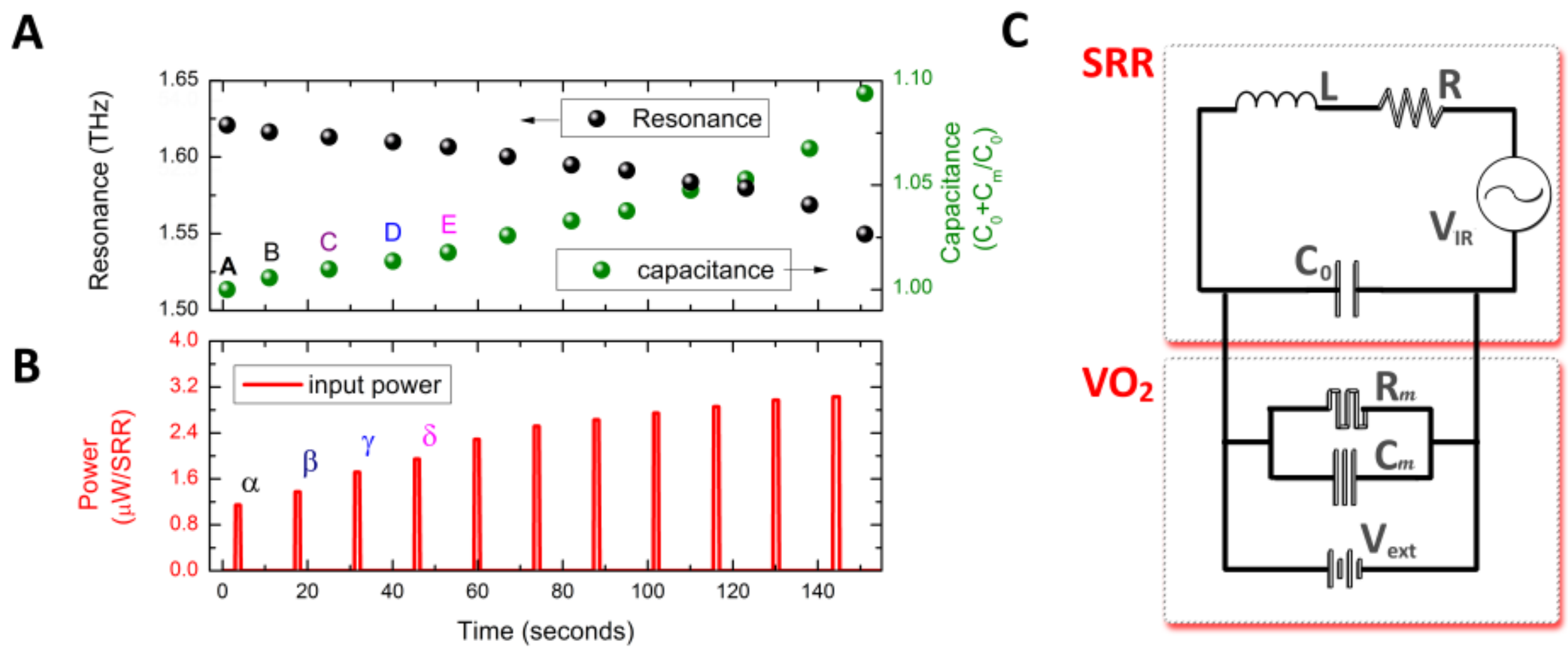

**Figure 2. Persistent electrical tuning of a metamaterial. A**, Successive modification of the resonance of the hybrid-metamaterial is achieved by sequential transient one-second electrical pulses of increasing power (**B**). These modifications persist until the device is thermally reset, and have been measured to be stable over 20min. This operation is well illustrated within the effective circuit model (**C**) by addition of a memory circuit elements $\mathbf{R}_m$ and $\mathbf{C}_m$ in parallel to the natural capacitance of the SRR.

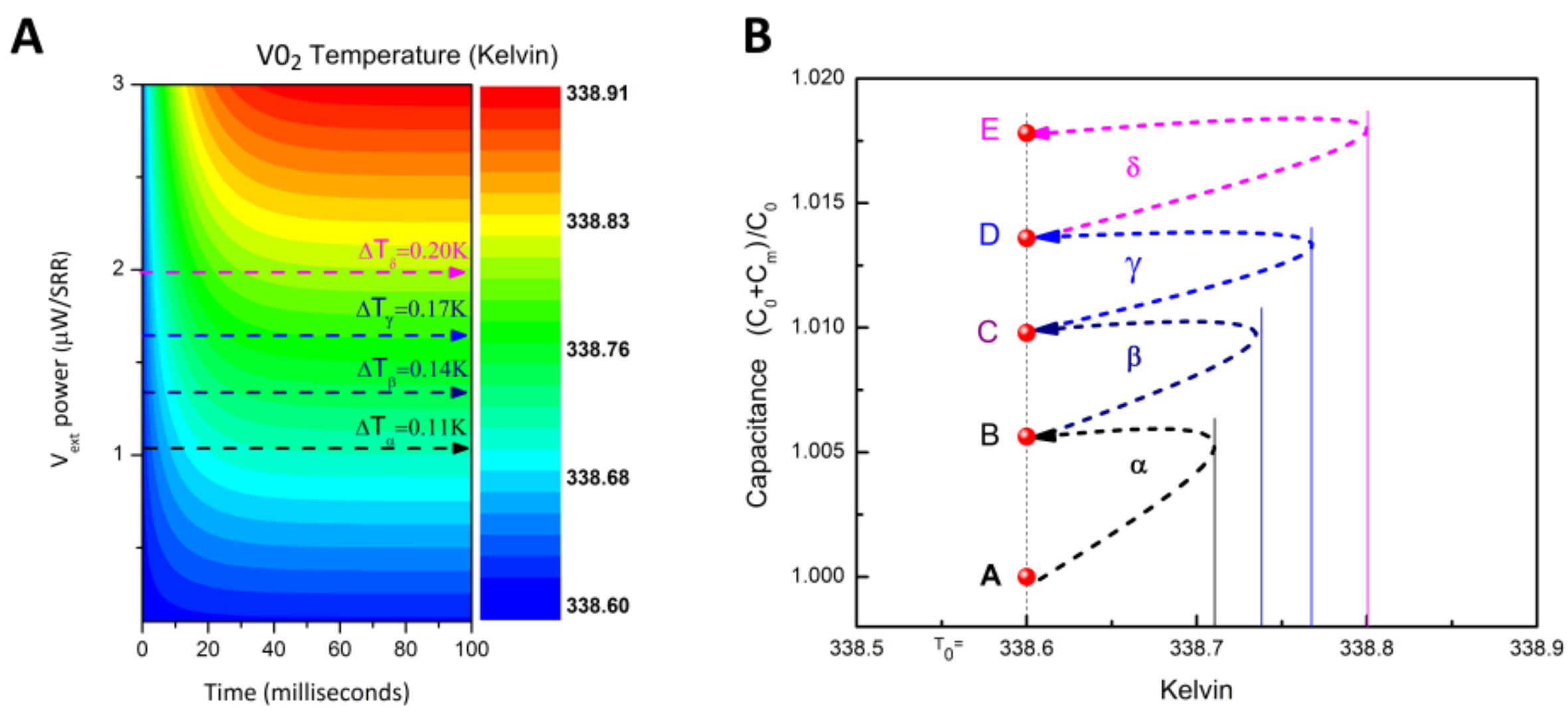

**Figure 3. The operation of memory capacitance explained via hysteretic phase transition. A**, Using a finite element model of our device, we calculate the temperature rise ΔT of the $VO_2$ film for a range of given input powers, including the first four voltage pulses in our experiment (α,β,γ,δ). A steady-state temperature rise is observed to emerge quite quickly. **B**, This temperature rise is used in combination with data from Fig.2A to estimate and sketch the behavior of capacitance during heating and cooling.

31. This work is supported by DOE, AFOSR and ETRI. MD acknowledges partial support from NSF. HK acknowledges research support from a project of Minister of Knowledge Economic in Korea.

# Memory metamaterials - supplementary information.

T.Driscoll, H.T. Kim, B.G. Chae, B.J. Kim, N. Marie Jokerst, S. Palit, D.R. Smith, M. Di Ventra, D.N. Basov

The following supplementary information is intended to complement the manuscript. It describes the experimental setup and procedure in more detail, including details of the thermal simulation used for Figure 3 of the main text.

## Materials and Methods

The vanadium dioxide is deposited on an $Al_2O_3$ substrate using the sol-gel method (S1). The gold metamaterial, a split ring resonator array, is photo-lithographically defined using a lift-off technique. Electrical leads are attached using silver epoxy, and the device is mounted on a temperature control stage inside a Bruker IFS-66v spectrometer. In this setup, we can simultaneously measure DC transport using a Keithley 6487 and the metamaterial Terahertz (THz) response. The metamaterial resonance frequency is identified as the minima in the sharp transmission dip. Spectra are recorded using the spectrometer's maximum resolution with a DFT zero-filling factor of 16. This allows us to identify our SRR resonance frequencies to within ~0.003 THz. Temperature control is done via Lakeshore 331 controller, with a measurement accuracy of 0.01 Kelvin. Tuning voltage pulses are applied using the Keithley and confirmed via oscilloscope.

## Thermal model.

The thermal modeling of our device is done in the commercially available COMSOL package for MATLAB (S2). A simplified 1-dimensional geometry consisting of 90nm $VO_2$, 500μm of $Al_2O_3$, and 3mm of Aluminum is created, depicted in Fig.S1. Input power is applied throughout the VO2 layer, simulating resistive heating. The heat-equation (Equation S1) is solved in the 3 regions as a Boundary Value Problem (BVP) using the finite element method. Continuity boundary conditions are used for the internal faces. The top boundary (above $VO_2$) is dT/dx = 0, approximating vacuum. The bottom boundary condition is T=338.6, approximating the large thermal mass of our heating stage combined with the PID feedback maintaining our set point temperature. The material thermal parameters used are listed in Table S1. This is repeated for a range of input power levels.

$$\rho c(T)\frac{\partial T}{\partial t} = \nabla \cdot (k \nabla T) = Q \qquad \text{S1}$$

Here, *T* is the temperature, $\rho$ is the density, *c(T)* is the specific heat at constant pressure, and *k* is the thermal conductivity.

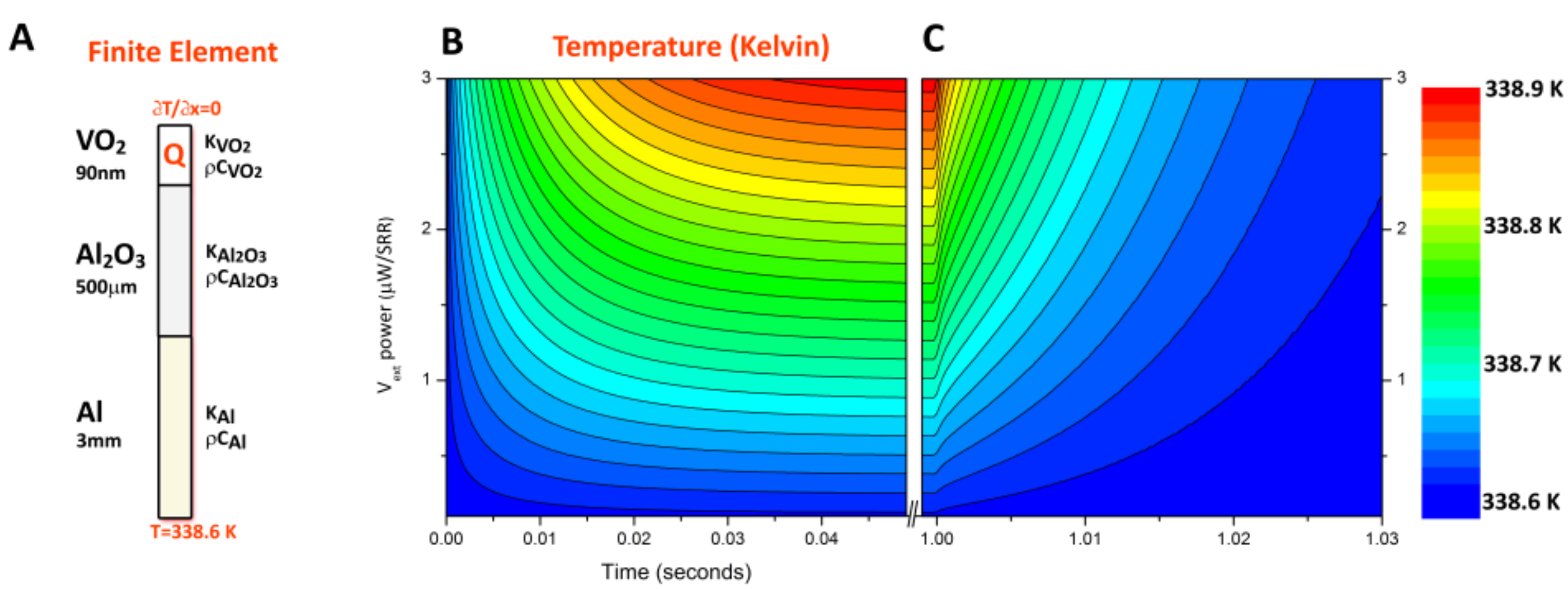

**Figure S1**. **A,** Finite element is used to solve the parabolic partial differential equation for our 3 layer system, giving $VO_2$ average temperature (**B**) in time for a span power levels. A nearly steady-state solution emerges within 50 milliseconds. Similarly, when the pulse stops after 1 second, the $VO_2$ layer quickly thermalizes (**C)** back to the bulk global temperature of 338.6 Kelvin.

Due to the presence of a phase-transition, the temperature dependant heat-capacity of $VO_2$ (**$c_{VO2}$(T)**) is difficult to accurately determine. Previous experimental data show a wide range of values for **$c_{VO2}$(T)** (S3-S5). In our model, we have added the insulating state heat capacity (843 J/Kg°K) and the known enthalpy of transition (*36*) averaged across a transition width of 4 Kelvin to arrive at a constant **$c_{VO2}$**. This simplification (as well as the assumption that other material parameters are not functions of temperature) reduces a nonlinear partial differential equation (PDE) to a simple parabolic PDE. Although this simplification for **$c_{VO2}$** is not mathematically rigorous, we find numerically that only the short-time (< 20ms) dynamics are affected by even large variations in **$c_{VO2}$(T)**. The steady-state solution, which emerges within ~50 milliseconds, is nearly insensitive to **$c_{VO2}$(T)**.

**Table S1**. Material values used for density (ρ), Heat capacity (**c**), and thermal conductivity (**K**) in the model. Parameters are for ~340 Kelvin.

| $\rho_{VO2}$ kg/m$^3$ | $C_{VO2}$ (J/Kg°K) | $K_{VO2}$ (W/m°K) | $\rho_{AL2O3}$ kg/m$^3$ | $C_{Al2O3}$ (J/Kg°K) | $K_{Al2O3}$ (W/m°K) | $\rho_{Al}$ kg/m$^3$ | $C_{Al}$ (J/Kg°K) | $K_{Al}$ (W/m°K) |
|---|---|---|---|---|---|---|---|---|
| 4340 | 1578 (S5) | 6.8 (S6) | 3938 (S7) | 752 (S7) | 30 (S7) | 2700 | 900 | 250 |